\documentclass[journal]{IEEEtran}
\usepackage{cite}

\ifCLASSINFOpdf
  \usepackage[pdftex]{graphicx}
  \graphicspath{{../fig/}{../jpeg/}{./}}
  \DeclareGraphicsExtensions{.pdf,.jpeg,.png}
\else
  \usepackage[dvips]{graphicx}
  \graphicspath{{../eps/}}
  \DeclareGraphicsExtensions{.eps}
\fi
\usepackage{amsmath}
\usepackage{amsfonts}
\usepackage{amssymb}
\usepackage{multirow}
\usepackage{multicol}

\usepackage{url}

\begin{document}
%
\title{Time-frequency Network for Robust Speaker Recognition}
%
%
%

\author{Jiguo Li,
        Xiaobin Liu,
        Lirong Zheng,~\IEEEmembership{Member,~IEEE,}
\thanks{Jiguo Li, and Lirong Zheng are with Micro nano System Center, the School of information science and technology, Fudan University, Beijing, 200433, China({email: jgli, lrzheng@fudan.edu.cn}). Xiaobin Liu is with Tecent.Inc, Beijing, 100193, China. (email: liu-xb@qq.com)}
\thanks{Correcording Author: lrzheng@fudan.edu.cn.}
}
%
%

\markboth{Journal of \LaTeX\ Class Files,~Vol.~14, No.~8, August~2015}%
{Shell \MakeLowercase{\textit{et al.}}: Bare Demo of IEEEtran.cls for IEEE Journals}
%



\maketitle
\begin{abstract}
 The wide deployment of speech-based biometric systems usually demands high-performance speaker recognition algorithms. However, most of the prior works for speaker recognition either process the speech in the frequency domain or time domain, which may produce suboptimal results because both time and frequency domains are important for speaker recognition. In this paper, we attempt to analyze the speech signal in both time and frequency domains and propose the time-frequency network~(TFN) for speaker recognition by extracting and fusing the features in the two domains. Based on the recent advance of deep neural networks, we propose a convolution neural network to encode the raw speech waveform and the frequency spectrum into domain-specific features, which are then fused and transformed into a classification feature space for speaker recognition. Experimental results on the publicly available datasets TIMIT and LibriSpeech show that our framework is effective to combine the information in the two domains and performs better than the state-of-the-art methods for speaker recognition.

\end{abstract}


%
\IEEEpeerreviewmaketitle

\section{Introduction}
\begin{figure}
   \centerline{\includegraphics[width=\columnwidth]{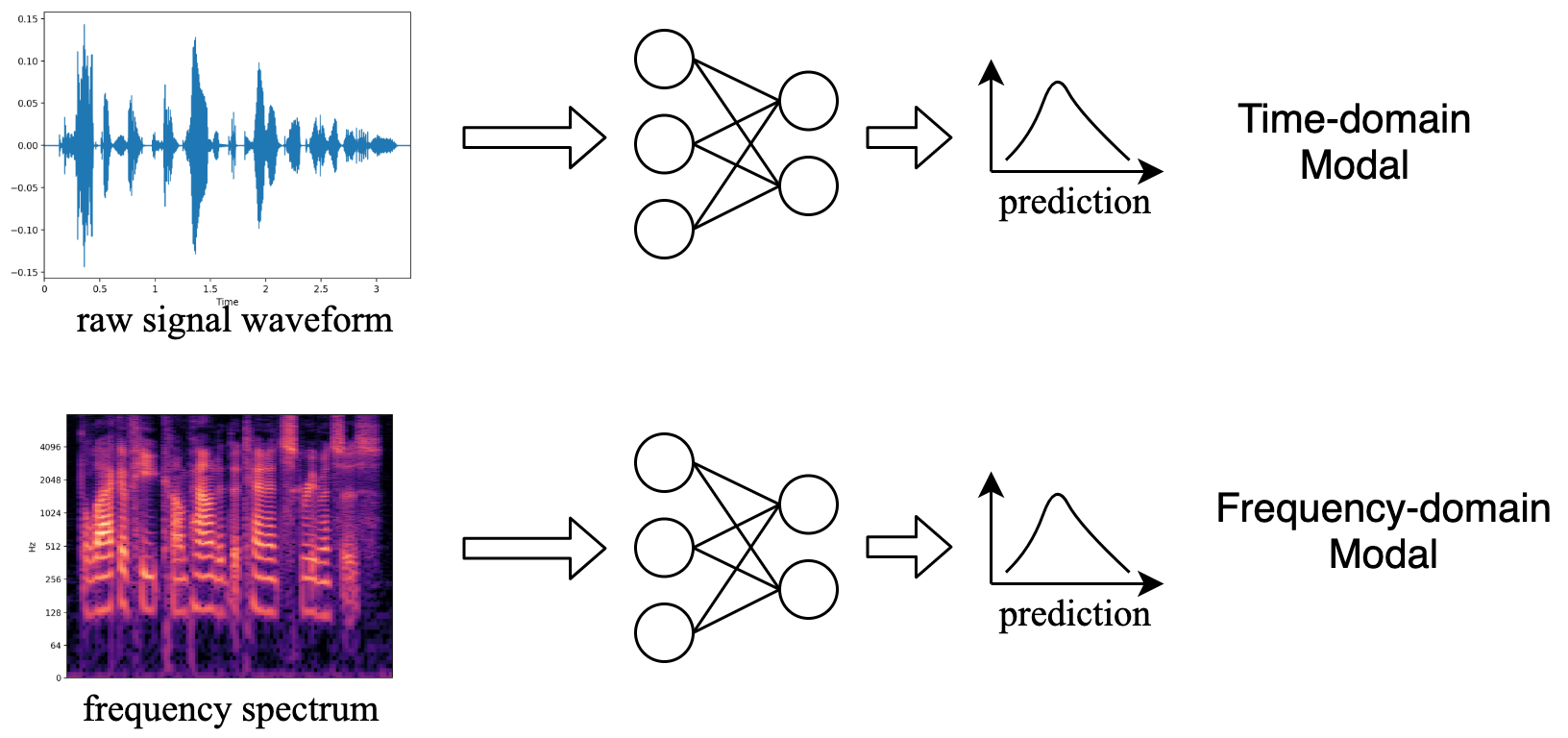}}
   \caption{The illustration of time-domain model~(upper) and frequency-domain model~(lower).~\textit{Time-domain model} uses a raw speech waveform as input.~\textit{Frequency-domain model} uses a frequency spectrum as input.}
   \label{fig:comparison}
\end{figure}

Speaker recognition is one of the most important areas in the speech processing community because of its wide applications in biometric authentication, speech recognition, forensics, and security. Most of the state-of-the-art speaker recognition methods can be divided into two categories according to the main domain of the input data: time-domain models and frequency-domain models, as shown in Fig.~\ref{fig:comparison}. The time-domain models use the raw speech waveform as input, which only has the time axis. The frequency-domain models adopt the frequency spectrum as input, which is a time-frequency representation of the speech signal. \textit{It is worthy noting that although the frequency spectrum also have the time information, the time resolution of the frequency spectrum is rather lower than that of the raw speech signal due to the window-based transformation~(such as STFT), so we classify these models which use the frequency spectrum as the input into the frequency-domain models.}
Based on time/frequency domain models which only use the information from single domain, we attempt to combine the information in both time and frequency domains and propose the time-frequency model, as illustrated in Fig.~\ref{fig:comparison}, which extracts and fuses features in both domains to leverage information as much as possible.
\\
\indent Before the deep neural networks~(DNNs) were applied in speaker recognition, most researchers used the frequency-domain features to classify the speech signals. Campbell~\textit{et al.}~\cite{campbell2006svm} modeled the speech signals with the Gaussian mixture model~(GMM).
Dehak~\textit{et al.}~\cite{dehak2010front} extracted the i-vector feature representation of speech segments to deal with speaker recognition. 
Both the above two popular methods are based on the hand-craft frequency-domain features, such as Filter bank~(FBANK) or Mel-frequency cepstral coefficients~(MFCC).
Besides, with the wide usage of DNNs in speech processing, DNNs are also employed to extract the frame-level features in the frequency domain for speaker recognition~\cite{yaman2012bottleneck}. All the above frequency-domain methods only process the signals in the frequency domain and ignore the information in the time domain.

\begin{figure*}
   \centerline{\includegraphics[width=.99\linewidth]{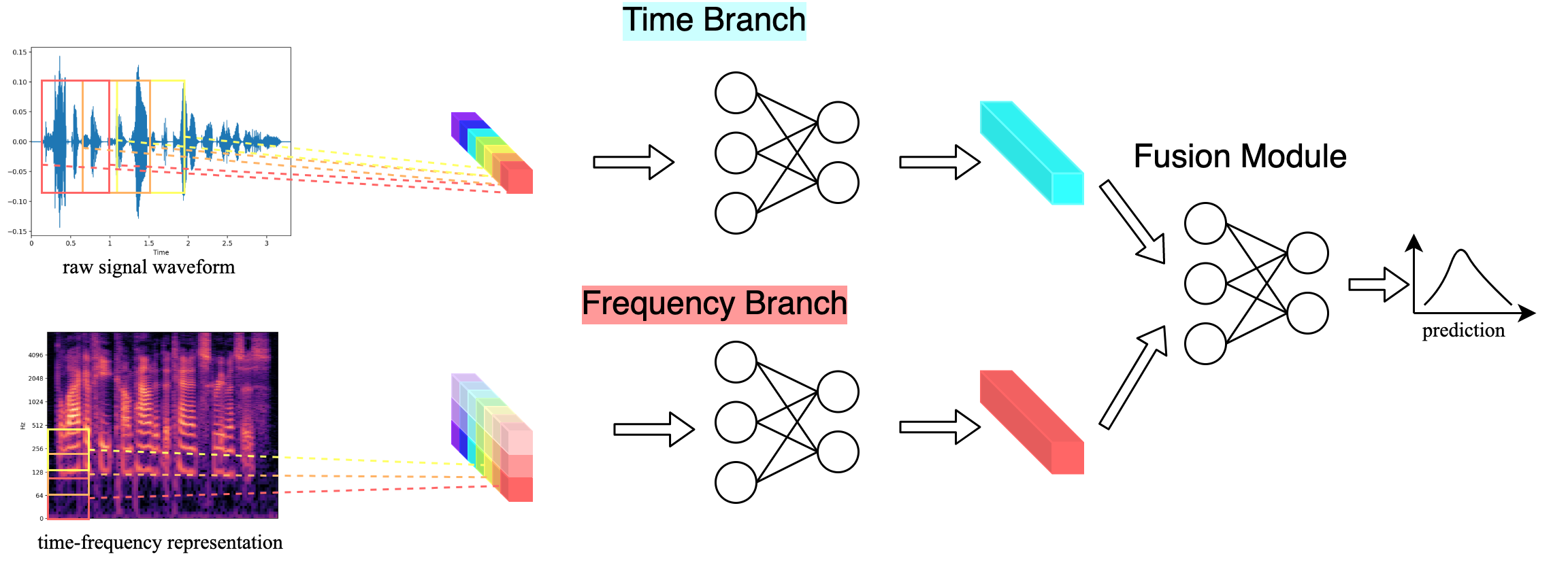}}
   \caption{Illustrations for the three submodules of our TFN. The time branch is based on SincNet with the first layer replaced with the learnable band pass filters to extract the time-domain features. The frequency branch is a multi-layer CNN to extract the frequency domain features. The fusion model is designed to fuse the two features from the time and frequency domains and outputs a global embedding to represent the speech signal.}
   \label{fig:three_submodules}
\end{figure*}


Researchers attempted to deal with speaker recognition directly in the time domain when convolution neural networks~(CNNs) were successfully applied in large scale images classification and showed powerful capability in modeling the high-dimensional data~\cite{krizhevsky2012imagenet,deng2012ilsvrc}. The end-to-end frameworks, which were designed to extract features from the raw speech waveform, were proposed in recent years~\cite{variani2014deep, heigold2016end} and achieved better performance than the traditional frequency-domain methods. However, these frameworks did not take the information of the frequency domain into account explicitly. Recently, SincNet~\cite{ravanelli2018speaker} was proposed for speaker recognition by designing the first layer's filters as learnable band pass filters. Taking the frequency characteristics of speech signals into consideration, SincNet achieved state-of-the-art performance, surpassing the prior end-to-end frameworks. However, SincNet only extracts the features from the raw speech waveform, without taking the frequency domain information into account. Hence it may result in suboptimal performance.

Motivated by the prior works on frequency and time domains, we propose the time-frequency network~(TFN) to combine the information in time and frequency domains~(as illustrated in Fig.~\ref{fig:comparison}). Our contributions can be summarized as follows:
\begin{enumerate}
   \item We proposed the TFN to combine the information in time and frequency domain by designing two branches~(time branch and frequency branch) to make full use of the information in speech signals. To the best of our knowledge, our work is the first one to combine time-frequency domains with both the raw speech signal and frequency spectrum as input for speaker recognition.
   \item The proposed TFN achieves state-of-the-art speaker recognition performance on large scale public datasets TIMIT and LibriSpeech.
   \item The ablation study about the fusion module indicates that more transformation layers for the fused feature can boost the recognition performance, which will help design the deep models with multi-inputs.
\end{enumerate}

The rest of this paper is organized as follows. Section~\ref{sec:related_works} shortly reviews the prior works about speaker recognition in frequency or time domain. Section~\ref{sec:tfn} introduces our proposed TFN. Section~\ref{sec:experiments} presented the experimental results of our proposed frameworks. Section~\ref{sec:conclusin} concludes the paper.

\section{Related Works}\label{sec:related_works}
\subsection{Speaker Recognition in Frequency Domain}
Before DNNs were employed in speaker recognition, researchers tended to design a hand-crafted global feature in frequency to represent the speech signal. Campbell~\textit{et al.}~\cite{campbell2006svm} combined the support vector machine~(SVM) with the GMM supervector and derived a linear kernel based on an approximation to KL divergence between two GMM models. Dehak~\textit{et al.}~\cite{dehak2010front} modeled both the speaker and channel variabilities and presented a new low-dimensional speech global representation, named identity vector or i-vector, which is the base of the most following frequency-domain methods for speaker recognition. Based on the i-vector, Lei~\textit{et al.}~\cite{lei2014novel} attempted to use a pretrained DNN rather than the stand GMM to produce the frame alignments and improved the equal error rate by 30\% compared with the baseline systems. Besides, x-vector~\cite{snyder2018x} was introduced by training a DNN on the frequency features~(such as FBANK) with the data augmentation to extract the global embeddings with a fixed length. Although DNNs are used in x-vector, it is still trained on the frequency features because the input of the DNNs is the frequency spectrum. 

\subsection{Speaker Recognition in Time Domain}
Researchers did not attempt to process the speech signals on the raw speech waveform, which only has the time axis, for speaker recognition until CNNs are used in this task~\cite{muckenhirn2018towards,jung2018complete}. Muckenhirn~\textit{et al.}~\cite{muckenhirn2018towards} firstly processed the speech signals with the raw speech waveform as input for speaker recognition, and only extracted the feature in the time domain with an end-to-end manner. Jung~\textit{et al.}~\cite{jung2018complete} combined the CNNs and the long short term memory~(LSTM) to extract the global embeddings for speaker recognition in an end-to-end manner from the raw speech signals, too.
Recently, another time-domain framework, SincNet~\cite{ravanelli2018speaker} was introduced by replacing the first layer of the CNNs with the learnable band pass filters to obtain better interoperability, achieving superior results to the plain CNN-based methods.
Although taking the frequency characteristics of the speech signals into consideration and using learnable band pass filters in the first layer, we still think that SincNet is a time-domain model because it uses the raw speech waveform as the input. SincNet is the baseline method due to its promising performance in speaker recognition.

\subsection{Jointly Time-frequency Learning}
Because time and frequency are the two most important domains for speech signal analysis~\cite{qian1999joint}, jointly time-frequency analysis for speech has been investigated for several decades~\cite{stankovic1994method, hess1996wavelets, qian1999joint}. When deep neural networks~(DNNs) showed its surpassing performance on feature representation learning for speech~\cite{hinton2012deep} and image~\cite{krizhevsky2012imagenet} data, researchers began to learn both time-domain and frequency-domain features with DNNs. T{\'o}th~\textit{et al.}~\cite{toth2014combining} proposed to conduct convolution along both time and frequency axes on a time-frequency representation~(such as Mel Bank Features) for phone recognition, resulting in a frequency subnetwork in a time network framework. Similarly, Mitra~\textit{et al.}~\cite{mitra2015time} proposed to extract time/frequency representation from the time-frequency acoustic features~(such as Normalized Modulation Coefficients) with two branches and concatenate them for the following speech recognition. Although these works attempted to combine the time and frequency information, they only used the frequency spectrum as input, which has a rather lower time resolution than the raw speech signal. Our proposed framework takes the time and frequency information and uses both the raw speech waveform and frequency spectrum as input by designing two branches to extract features for the time and frequency domains.
 As far as we know, our work is the first one to combine time-frequency domains with two branches for speaker recognition.

\section{Time-frequency Network}\label{sec:tfn}
As illustrated in Fig.~\ref{fig:three_submodules}, our TFN contains two branches~(a time branch and a frequency branch) and a fusion module. The time branch is based on the SincNet~\cite{ravanelli2018speaker} due to its promising results on speaker recognition. The frequency branch is a multi-layer CNN with a frequency spectrum representation~(such as MFCC) as input and the fusion module is designed to fuse the time-domain feature and the frequency-domain feature.~\textit{It is worth noting that although the frequency spectrum also contains the time information, its time resolution is much lower than the raw signal due to the window-based transformation~(such as STFT).} 
\subsection{Time Branch}
The time branch is designed to extract the time-domain feature from the raw speech signals. Motivated by the SincNet's~\cite{ravanelli2018speaker} promising performance and interpretability, the time branch of our framework is designed based on SincNet. Following SincNet, the first layer of the time branch is designed as the learnable band pass filters to model the frequency characteristics for speaker recognition. A band pass filter can be formulated as:
\begin{equation}
   \begin{aligned}
   g[n, f_1, f_2 ] = 2f_2\text{sinc}(2\pi f_2 n)-2f_2\text{sinc}(2\pi f_1 n),
   \end{aligned}
\end{equation}
where $f_1/f_2$ is the low/high cutoff frequency, $\text{sinc}(x)=\sin{x}/x$. By designing the filters as band pass filters, the model has fewer parameters and better interpretability. Except for the first layer, all other layers in our time branch are the typical 1-dimensional convolutional layer~(Conv), followed by the batch normalization layer~(BN)~\cite{ioffe2015batch} and ReLU layer~(ReLU)~\cite{he2015delving}. After several convolutional blocks~(Conv, BN, and ReLU), we get a time-domain feature representation. Although the first layer of SincNet is the learnable band pass filters, we regard it as a time-domain model because it is with the raw speech signals as input, which only has the time axis. 

\subsection{Frequency Branch}
The frequency branch is used to extract the frequency-domain feature from the frequency spectrum. As illustrated in Fig.~\ref{fig:three_submodules}, the frequency spectrum can be obtained by imposing a time-frequency transformation~(such as STFT) onto the raw speech signals. The spectrum can be seen as a feature map and we can leverage CNNs to extract the frequency feature, referring the prior frequency-domain works~\cite{lei2014novel, snyder2018x}. Several sequential convolutional blocks~(Conv, BN, and ReLU) are used to learning the high-level representation before the frequency-domain embedding is fed into the fusion module.
\subsection{Fusion}
\begin{figure}
   \centerline{\includegraphics[width=.85\linewidth]{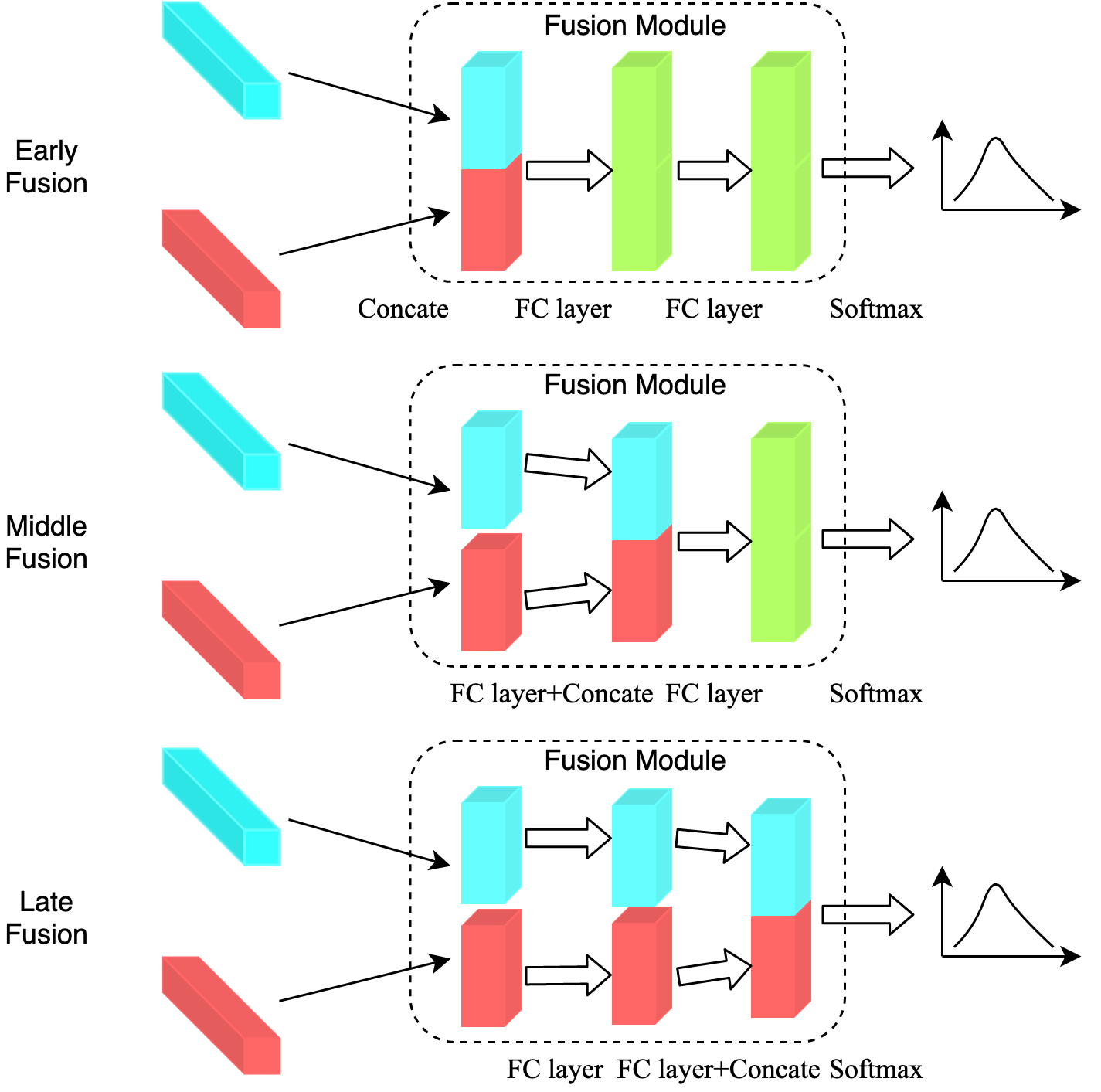}}
   \caption{Three different fusion types: early fusion, middle fusion and late fusion.}
   \label{fig:fusion_types}
\end{figure}

The fusion module is designed to fuse the two domain's features by concatenating the two features and transforming the global feature into a classification feature space. According to the different transformation types, the fusion module has three different implementations: early fusion, middle fusion, and late fusion. As illustrated in Fig.~\ref{fig:fusion_types}, the early fusion inputs the local embeddings and uses a two-layer transformation to project the concatenated global feature into a classification feature space, while the middle fusion transforms the two domain-specific features respectively and then projects the concatenated global feature into the classification feature space with only one transformed layer. The late fusion has no global transformation layer and only concatenates the two domain-specific features after two intra-domain transformation layers. To investigate which fusion method is the best, we conduct an ablation study and the results are shown in Section~\ref{sec:experiments}.
\subsection{Implementation Details}
Following the experiment setting in SincNet~\footnote{https://github.com/mravanelli/SincNet/}, the band pass filter number is set as 512/1024 for the small/big models for both TIMIT and LibriSpeech. 
Besides, we use MFCC as the frequency spectrum in our frequency branch because MFCC has been widely used in speaker recognition. The model is trained end-to-end and Adam~\cite{kingma2014adam} is used to train the model with a learning rate $1e-3$.
We follow the SincNet\cite{ravanelli2018speaker} for the model initialization and the data preparation. All the models are trained for 24 epochs, following the setting in~\cite{ravanelli2018speaker}.
\section{Experiments}\label{sec:experiments}
\subsection{Datasets and Metrics}
Following SincNet~\cite{ravanelli2018speaker}, we conduct experiments on TIMIT~\cite{garofolo1993darpa}~(462 speakers, train chunk) and LibriSpeech~\cite{panayotov2015librispeech}~(2484 speakers, train chunk) to show the effectiveness of our proposed method. 
The training/testing split follows the official implementation
of SincNet~\cite{ravanelli2018speaker}, in which 2310/1386 samples are used for training/testing in TIMIT, and 14481/7452 samples are used for training/testing in LibriSpeech. The classification error rate~(CER) is used to evaluate the performance of the proposed and baseline models.

\subsection{Comparison with Baselines}
\begin{table}
\centering
\caption{Experimental results on TIMIT and LibriSpeech datasets. }
\label{tab:experimental_result}
\begin{tabular}{c|c|c|c}
\hline
Dataset & Model Type & Model Size~(Mb) & CER~(\%)$\downarrow$ \\
\hline
\multirow{7}{*}{TIMIT} & SincNet~\cite{ravanelli2018speaker} & 78 & 0.85 \\
\cline{2-4}
                    & TFN~(freq only, small) & 11 & 1.01 \\
                    & TFN~(time only, small) & 63 & 3.18 \\
                    & TFN~(freq only, big)   & 41 & 0.72 \\
                    & TFN~(time only, big)   & 73 & 3.03 \\
                    & TFN~(both, small)      & 73 & 1.73 \\
                    & TFN~(both, big)        & 118 & \textbf{0.65} \\
\hline
\multirow{7}{*}{LibriSpeech} & SincNet~\cite{ravanelli2018speaker} & 103 & 0.96 \\
\cline{2-4}
                    & TFN~(freq only, small) & 15 & 0.34  \\
                    & TFN~(time only, small) & 67 & 0.52  \\
                    & TFN~(freq only, big)   & 48 & 0.50 \\
                    & TFN~(time only, big)   & 86 & 0.67 \\
                    & TFN~(both, small)      & 81 & \textbf{0.31} \\
                    & TFN~(both, big)        & 134 & 0.32  \\
\hline
\end{tabular}
\end{table}

To demonstrate the effectiveness of our proposed TFN, we conduct comparison experiments with different model sizes by controlling the dimension of the classification feature space. For each dataset, we design a small network and a big network. For the TIMIT dataset, the small/big model's classification feature space dimension is 512/1024. For the LibriSpeech dataset, the small/big model's classification feature space dimension is 1024/2048. As illustrated in Table~\ref{tab:experimental_result}, two conclusions can be drawn from the results:
\begin{enumerate}
   \item \textit{Our TFN model performs better than the baseline model.} As shown in Table~\ref{tab:experimental_result}, for TIMIT dataset, our TFN~(both, big) can achieve a CER of 0.65\%, which is better than the baseline~(0.85\%). Similar results can be achieved on LibriSpeech Dataset. Our TFN~(both, small) shows better performance than the baseline~(0.31\% vs 0.96\%). Hence, we can conclude that our TFN model, which contains both the time and frequency branches surpasses the baseline, which is one of the state-of-the-art models for speaker recognition.
   \item \textit{Our two-branch TFN model surpasses the one-branch models mostly.} Another concern about our TFN model is whether the speaker recognition model can benefit more from the two-branch framework than that of the one-branch framework. As shown in Table~\ref{tab:experimental_result}, Most of the TFN models with both branches show better performance than the single branch model. For example, On TIMIT dataset, TFN~(both, big) is superior to TFN~(freq, big) or TFN~(time, big). On the LibriSpeech dataset, TFN~(both, small) surpasses TFN~(freq, small) or TFN~(time, small). The only exception is that TFN~(both, small)'s performance is not as good as TFN~(freq, small). So in most cases, the speaker recognition model can benefit from the two-branch framework~(one for time domain and the other for frequency domain).
\end{enumerate}

\subsection{Ablation Study for Fusion Type}
\begin{table}
   \centering
   \caption{Ablation study on different fusion types on TIMIT and LibriSpeech datasets}
   \label{tab:ablation_study}
   \begin{tabular}{c|c|c|c}
   \hline
   Dataset & Fusion Type & Model Size~(Mb) & CER~(\%)$\downarrow$ \\
   \hline
   \multirow{3}{*}{TIMIT} & Early Fusion & 34 & \textbf{1.15} \\
                       & Normal Fusion & 32 & 1.44 \\
                       & Late Fusion   & 30 & 1.87 \\
   \hline
   \multirow{3}{*}{LibriSpeech} & Early Fusion & 42 & \textbf{0.52} \\
                       & Normal Fusion & 40 & 0.53 \\
                       & Late Fusion   & 38 & 0.54 \\
   \hline
   \end{tabular}
   \end{table}

   We conduct an ablation study to compare the speaker recognition performance of different fusion types.
   As shown in Table.~\ref{tab:ablation_study}, on both TIMIT and LibriSpeech Datasets, the model with early fusion performs best among the three fusion types, indicating more inter-domain transformation layers for the global feature~(concatenating the time-domain feature and frequency-domain feature) can boost the recognition performance. The reason may be more transformation layers result in more learnable parameters, so the model can fit the data better. However, more inter-domain transformation layers may result in overfitting if the training data are not enough for training the model.

\section{Conclusion}\label{sec:conclusin}
In this paper, we proposed the time-frequency network~(TFN) for speaker recognition to combine the time-domain information and frequency-domain information by designing two branches for learning time and frequency domain feature representation. The time branch uses raw speech waveforms as input, while the frequency branch adopts the frequency spectrum as input, which has a lower time resolution than the raw signal. Experimental results on TIMIT and LibriSpeech dataset showed that our proposed TFN surpassed the state-of-the-art baseline and the two-branch framework performs better than the one-branch framework, demonstrating the effectiveness of our proposed TFN model. The ablation study about the fusion module showed that more inter-domain transformation layers for the concatenated features can boost the recognition performance.

\ifCLASSOPTIONcaptionsoff
  \newpage
\fi



%
\bibliographystyle{IEEEtran}
\bibliography{refbib}

%








\end{document}